# The Benefits of Edge Computing in Healthcare, Smart Cities, and IoT

**Rushit Dave**[*], **Naeem Seliya, Nyle Siddiqui**

Department of Computer Science, University of Wisconsin – Eau Claire, Eau Claire, US
*Corresponding author: daver@uwec.edu

**Abstract**  Recent advancements in technology now allow for the generation of massive quantities of data. There is a growing need to transmit this data faster and more securely such that it cannot be accessed by malicious individuals. Edge computing has emerged in previous research as a method capable of improving data transmission times and security before the data ends up in the cloud. Edge computing has an impressive transmission speed based on fifth generation (5G) communication which transmits data with low latency and high bandwidth. While edge computing is sufficient to extract important features from the raw data to prevent large amounts of data requiring excessive bandwidth to be transmitted, cloud computing is used for the computational processes required for developing algorithms and modeling the data. Edge computing also improves the quality of the user experience by saving time and integrating quality of life (QoL) features. QoL features are important for the healthcare sector by helping to provide real-time feedback of data produced by healthcare devices back to patients for a faster recovery. Edge computing has better energy efficiency, can reduce the electricity cost, and in turn help people reduce their living expenses. This paper will take a detailed look into edge computing applications around Internet of Things (IoT) devices, smart city infrastructure, and benefits to healthcare.

## 1. Introduction

Technology has been making its way into every facet of our lives. There are many different smart devices, sensors, and even self-driving cars; however, cloud computing-based processing for such devices and machines has been relatively inadequate. Edge computing could be an effective means of improving the performance of such devices and machines, thereby addressing the short comings of cloud computing [1,2,3]. Edge computing is an architecture that integrates computing, storage, and network services that extend from cloud computing onto the network edge. Compared to cloud computing services, edge computing improves speed of computing, minimizes storage, lowers latency bandwidth, improves security of data, and reduces location limitation by offloading certain computations to edge devices.

Within edge computing, massive amounts of data will be generated by various types of devices. Instead of transmitting the data back to cloud services, the data can be computed in the edge service to reduce the cost of bandwidth and energy consumption [4]. Additionally, edge computing can also extract and sort the important data from the received raw data to reduce the amount of storage required. Moreover, edge computing provides efficient, secure services for many end-users. Edge computing has the advantage over cloud computing in the context of fifth generation (5G) mobile communication systems by utilizing the faster bandwidth and lower latency to ensure the data is processed and analyzed rapidly and efficiently.

In recent years, the idea of smart cities has become a trending concept for a new way of life. Smart cities implement many IoT devices to build an advanced, interconnected infrastructure thereby providing a convenient environment for living. In a typical smart city system, there are many facilities and applications, such as smart parking, smart buildings, smart lighting, advanced waste disposal, and traffic management [5]. Edge computing helps smart cities to achieve optimal energy management by ubiquitous monitoring. Furthermore, short-term energy consumption monitoring, analysis, and prediction provide data to aid in energy planning, distribution, and conservation. Consequently, this process can help to reduce energy usage and monetary waste on electricity, bandwidth, and save on personal time. With an increase in city growth, additional resources are required which lower latency in the system; thus, demanding a higher bandwidth need. Therefore, the excellent scalability feature of edge computing provides it an advantage over regular cloud computing.

There is a rapidly growing amount of IoT devices which produce data and need computation by means of cloud computing, especially when machine and deep learning algorithms are used as predictive models [6,7,8]. Due to the certain situational circumstances, the offloading bandwidth for the application is low which means it has a high latency. With this heavily increased interest in IoT technology, cloud computing has become bottlenecked by latency, high-cost maintenance, and service. Edge computing can easily solve the disadvantages of cloud computing. With the 5G bandwidth speed, data from IoT devices can be processed efficiently by edge computing due to its low latency [9]. Even though IoT devices have made great improvements recently, several constraints are still present such as battery life, memory size, and cooling systems. Edge computing provides solutions to these constraints, and in turn, maintains and extends the devices' functional lifetime.

In addition to smart city technology and IoT applications, the healthcare domain is another critical domain that can benefit from edge computing. Edge computing can reduce the amount of data in circulation and improve efficiency within the healthcare domain [10]. Furthermore, edge computing architecture can also help medical staff reduce the dependence on remote centralized servers [11,12] and increase data security and ethical integrity [13,14,15]. Currently, wearable devices and sensors are utilized as a treatment and active monitoring option for at-home patients who suffer from Parkinson's disease, a high risk of heart attacks, and other severe ailments [16]. Edge computing can react fast with low latency in conjunction with these devices to increase reliability and to prevent an untoward incident in these scenarios.

In this paper, we will discuss how edge computing has positively affected the domain areas of the Internet of Things (IoT), smart cities, and healthcare, and how they can continue to do so. These three imperative categories are chosen due to our growing dependence on them in our everyday life. This paper will provide technical reviews of how the most recent literature has exploited edge computing to improve performance among other aspects in each of the aforementioned domains. These categories are essential for building a sustainable future and thus it is essential that high quality reliability and user experience is a top priority. Edge computing is not without its current limitations [17] and the security risks that IoT devices can pose is one of the biggest concerns for users, hence why further research in edge computing systems is necessary. This paper aims to highlight these limitations to be improved upon in the future. Edge computing allows for more secure and efficient transfer of data when compared to cloud computing, making edge computing a viable solution to the multitude of security vulnerabilities that currently plague cloud computing. Since raw data is processed and computed on an edge computing platform, vulnerable raw data is not returned to the cloud or database, minimizing the probability of a successful attack and lowering the risks of data transmission.

## 2. Background

Edge computing is one of the latest technologies to ensure that our data will have an improvement in quality, security, and processing time. The most important of these improvements is the reduction of latency in products which are time-critical [2,18]. For example, edge computing can effectively address the problem of driverless cars possessing an essential need for near real-time responses to incoming data. Edge computing can also help a device extend its battery life and lower its battery usage by computing the data on the edge as opposed to locally on the device where data need not be transferred to the cloud. Such a strategy can reduce bandwidth and in turn improve battery life. Due to the introduction and advancement of 5G networks, they have much higher speeds than 4G networks and are more compatible with edge computing systems. 5G will help the edge computing networks by lowering the latency between multiple devices.

IoT consists of physical devices that are connected to the internet, and which are collecting and sharing data in the context of a particular system [19]. The Apple Watch, other fitness brand devices, remote lights, etc. are all examples of IoT devices. All these devices are connected to the internet and produce data that is sent back to their company or to a database server to be recorded, collected, and analyzed. Since IoT systems mandate a rapid system response from/to connected devices, edge computing is more compatible with the requirements of IoT systems than cloud computing. The process of edge computing has the advantages of low latency and reducing battery consumption when used in conjunction with IoT devices. As concluded in [20], increased utilization of edge computing will improve the Quality of Service (QoS) and the overall user experience of IoT devices. Due to the abundance of data being generated in this day in age, continuing to rely on cloud computing with the current cost of bandwidth will be a big challenge for the average consumer with many different IoT devices [21]. Moreover, increased number of data transmissions increases the risks of malicious data leaks and potential security susceptibility. To this extent, [22] exhibited how computing data on the edge reduces the risk of data security breaches.

Smart cities are cities that use advanced technology and sensors alongside artificial intelligence algorithms to improve the human living experience [23-25]. Their aim is to make everyday life more convenient and streamline and automate time consuming day-to-day tasks [26, 27]. Authors of [28] highlighted many examples of these benefits, such as a rice cooker starting to cook rice when a user commutes back from work, the heater or air conditioner activating before a user is home, CCTV cameras accurately identifying and reporting suspicious individuals walking around residences, and more. They further assert that edge computing is a necessary tool in addition to the previous examples to help people monitor their devices at home and ensure better management of energy consumption; this prevention of wasted electricity leads to decreased pollution and reduces the exploitation of increasingly scarce resources. A smart city system already aids in heavily reducing emissions and pollution, so incorporating edge computing into this system will only bolster this effect. Since the price of electricity is a growing concern for this generation because of the number of electric devices in circulation as seen in [29], it further proves the necessity of edge computing.

The important application of edge computing within the healthcare domain is present in smart healthcare frameworks such as medical resource management, medical devices, and 3D remote computed tomography (CT) imaging [30] and healthcare automation [31]. [30] states the need for edge computing in medical management is to equip hospitals with higher quality methods of patient interaction as well as providing a more seamless way to conduct visits and submit prescriptions. Additionally, medical devices like electrocardiogram (ECG) devices seen in [32] require an instant response for proper and efficient use; a functional aspect that the low latency times of edge computing can support. This advantage of low latency on edge computing can also apply to remote 3D CT images. CT images are widely used on patients to diagnose clinical diseases and identify important characteristics for pathological diseases. [20] exhibits how edge computing can be used with medical devices to evaluate patients, collect data, and transmit the images back to the doctor in real-time to reduce the diagnosis window and start treatment right away. Other state-of-the-art technologies have also been applied to healthcare and other domains to improve the quality and efficiency of service [33,34].

## 3. Literature Review

### 3.1. Novel Edge Computing Techniques for IoT

The authors of [35] discuss edge computing technologies with relation to IoT. They propose a combination of edge computing and cloud computing (called cloudlet) for computing on IoT devices and explain the challenges that must be overcome. Mobile Edge Computing (MEC) is the new technology defined by European Telecommunications Standards Institute (ESTI) which provides supporting services to informational technology processes and is compatible with cloud computing at the edge of the mobile network. The novel innovations of MEC are in the computational offloading and mobility management methods. [35] compares and contrasts fog computing versus edge computing. As opposed to edge computing, fog computing operates with additional overhead management in the computational process by providing tools for distributing, orchestrating, managing, and securing resources between devices residing on the edge. The cloudlet is the solution for cloud computing and end-to-end responsiveness between an IoT device and the cloud system. The main objective of cloudlet is to provide computing resources to mobile devices with lower latency by accessing nearby cloudlet services.

Despite the success of edge computing in improving the service around IoT, it continues to face new challenges in this field. One challenge being big data mining due to the mass amounts of data being transferred from IoT devices to an edge computing platform. Edge computing must define what is the important and relevant data to compute before sending that data to the cloud. Another challenge edge computing faces is network slicing. Network slicing is a network architecture that enables multiplexing of virtualized and independent logical networks on the same physical network infrastructure which may not be able to fully realize the advantages of edge computing technology due to its parallelism.

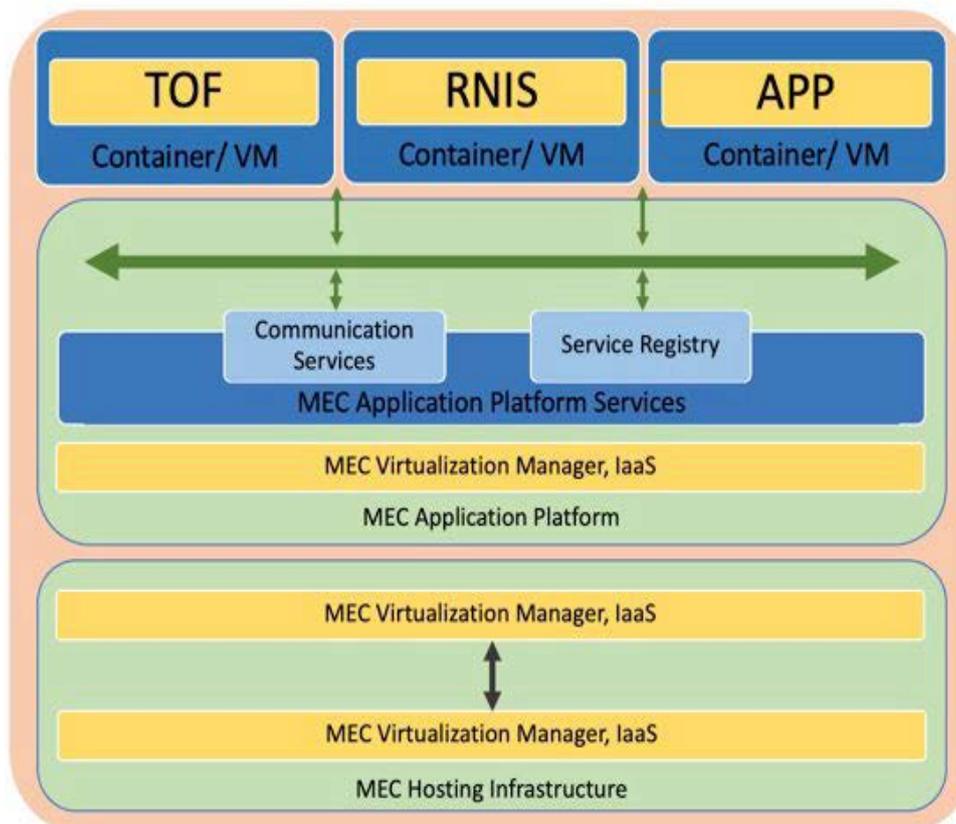

**Figure 1.** Proposed ETSI-compliant vMEC architecture

[36] further explores the aforementioned dependency of 5G networks that edge computing solutions possess and proposes the 5G virtualized Multiaccess Edge Computing platform (vMEC) architecture for IoT applications. The vMEC architecture is comprised of three parts: the host infrastructure, the application platform, and application services shown in Figure 1. vMEC's architecture revolves around disaggregating hardware devices into actual, physical hardware devices and an associated virtualization layer which provides computing power, data storage, and control functions. The architecture uses two applications that call Traffic Offloading Function (TOF) and Radio Network Information Service (RNIS) to speed up the response time of flow control with poor wireless signal strength. TOF is tasked with traffic priority judgement and path selection based on current data flow and network traffic. RNIS records the underlying radio network information between base stations and users to further improve policy decision time. Moreover, the architecture contains an intermediary layer of middleware applications and software to aid in network traffic control between the edge and cloud and other communication services that may require intermediary software registration. These architectures were implemented with the two main incentives of ultra-low latency and high-availability for IoT application platforms. Various service configurations were tested to simulate different applicational situations and the authors concluded that a platform such as vMEC reduced latency of IoT application services by an average of 30%. They hypothesize that this is due to the superior flexibility and deployability vMEC possess over conventional network traffic control strategies.

The harmonic execution of Software Defined Networking (SDN), Network Function Virtualization (NFV), and vMEC architecture are all imperative in the potential generation of a low latency and high bandwidth environment for the trillions of IoT devices in use around the world.

[37] proposes the use of machine and deep learning to apply edge computing in an intelligent iRobot-Factory to increase the efficiency of production, as seen in the Figure 2. The cloud computing layer includes both the data center and cognitive engine systems to provide the ability for large number of edge users and devices to operate simultaneously. It then follows that the architecture of an intelligent robot must also have other intelligent devices that utilize cloud and edge computing congruently to relieve congestion within the system. The role of edge computing in this data domain is to implement better data security and furthermore caching and transmitting important data to upload to the cloud.

In addition, edge computing in an iRobot-Factory can assist the robot sensors in accurately analyzing external stimulus and production data in real-time. The authors further state that the dual implementation of edge computing and cloud computing in an iRobot-Factory also provide a viable method to increase failure prediction, model confidence estimation, and real-time cloud computing feedback; thereby decreasing instruction issues for operators and management.

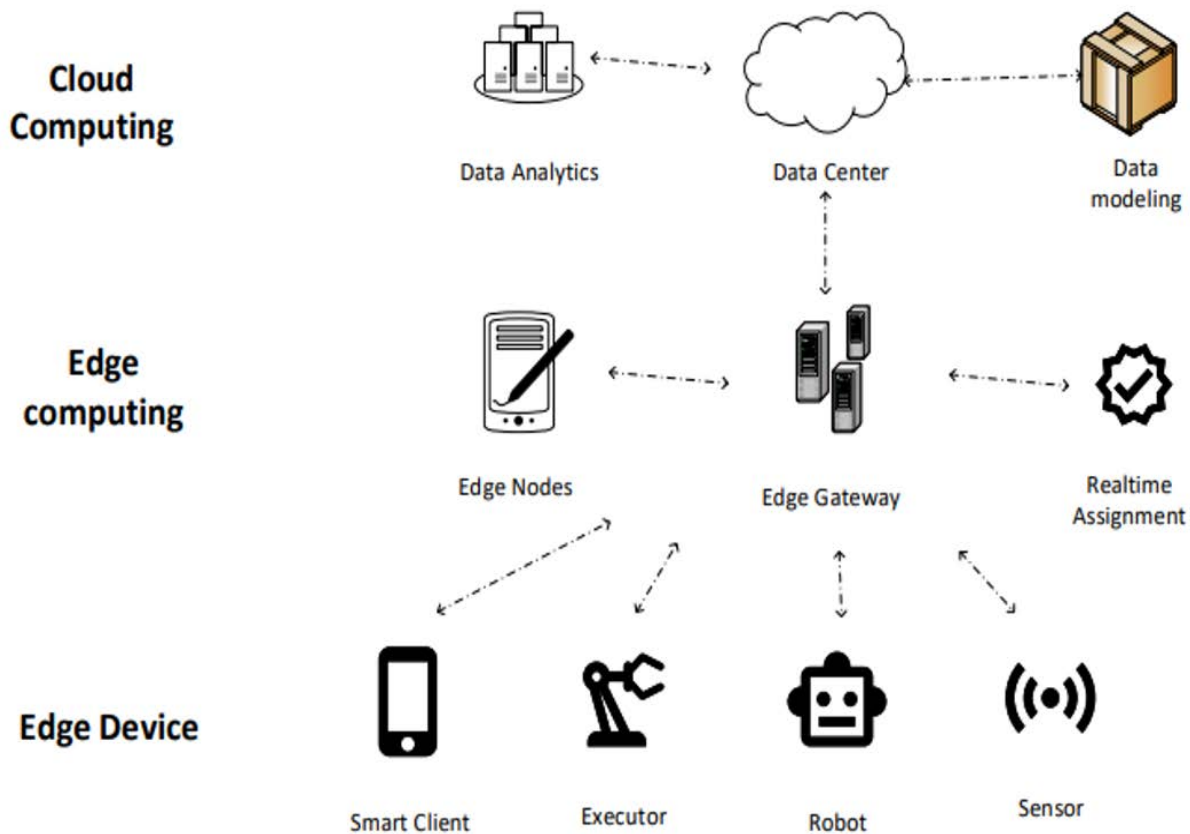

**Figure 2**. Intelligent communication based on edge computing. Taken from [37]

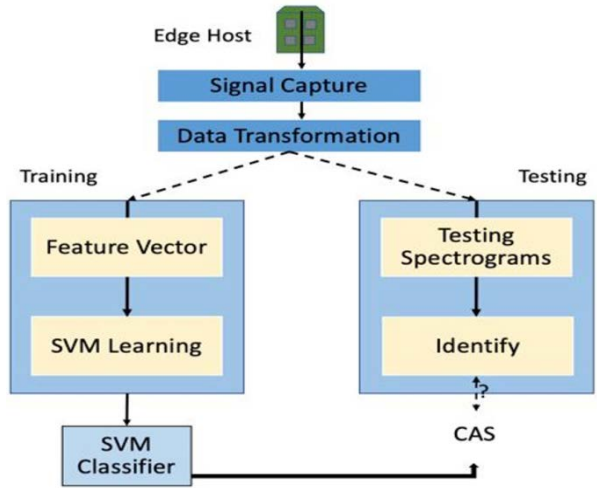

**Figure 3**. The frequency distribution of a 5V charger under different loads from [38]

Wang et. al in [38] attempt use electromagnetic radiation (EMR) from heterogenous devices as a continuous authentication scheme for edge computing. They hypothesize that the EMR signals produced by IoT devices during the edge computing process are distinct enough to classify each device. EMR is a signal that is formed by radiant energy produced by certain electromagnetic processes. These signals can be captured and decoded by an oscilloscope with a simple copper coil. Figure 3 exhibits the process for capturing these signals from an IoT device and using them to make an authentication decision. This paper finds that support vector machines (SVM) achieved upwards of 95% accuracy rates when using EMR as a unique authentication feature between edge and cloud computing devices to protect against illegitimate access. Additionally, an experiment was conducted and found that an authentication method consisting of a combination of edge computing and elliptic curve cryptography (ECC) methods could be effectively compromised. The experiment used a simple coupling coil to capture the wave signal of electromagnetic radiation from the switched-mode power supply (SMPS) and record data by the virtual oscilloscope.

They conduct the experiment by imitating a legitimate IoT device in the edge computing network and test the accuracies of various machine learning algorithms. The authors conclude that it is impossible to perfectly mimic the EMR signal from a specific device because the signal is produced by SMPS in an independent operating system in a safe computer room. Even though nefarious users have the ability to sniff data packages, they are still not able to decrypt them due to the strength of the ECC algorithm. The proposed method successfully uses EMR to fix potential security vulnerabilities within edge computing, such as spoofing, tampering, and replay attacks.

## 3.2. Novel Edge Computing Techniques for Smart Cities

In conjunction with the IoT domain, edge computing technology also plays an important role in smart cities. In a smart city, there is an amalgamation of IoT devices with sensors that generate massive amounts of data which require high bandwidth and low latency for instant responses. [39] describes an edge-based platform for a dynamic smart city application. The author of this paper proposes a novel application for smart cities called iSapiens that provides useful operational features for smart environments and cyber physical systems (Figure 4). The figure describes the operation for each component and feature for the installation and execution. They have implemented the proposed application into a "smart street" that extends for approximately 2.5 kilometers located in the Italian city of Cosenza, which has a population of around 70,000.

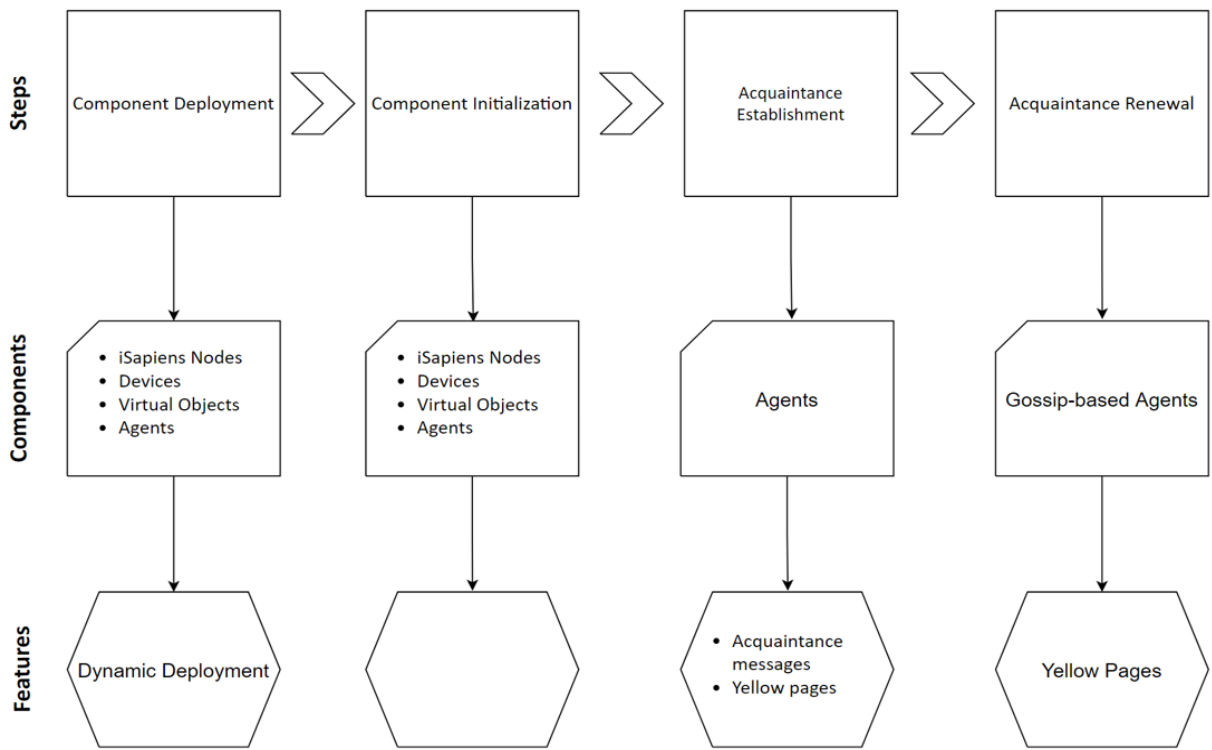

**Figure 4**. Operational steps for components installation and execution [39]

The smart street targets areas frequented often by pedestrian and general traffic allowing for collection of large amounts of robust and multi-faceted data. One previously overlooked feature found in this experiment was the attention to the humidity index for the city. Sensors allowing for the 24-hour a day collection of humidity can appropriately regulate temperature and air quality alongside humidity in a controlled smart city. Thus, the paper proposes iSapiens as a viable application to manage and control these types of features in smart cities using edge computing technology's strength in, system extensibility, fault tolerance, and vertical integration of all system processes.

Another key aspect of an efficient smart city is the optimization of energy consumption; another aspect of future smart city design that edge computing excels in. Authors of [40] developed a short-term energy prediction system (Figure 5) built on edge computing technology. The framework for this system is presented as a 4-stage hierarchal workflow: data acquisition and fusion, event data generation, scenario model establishment and prediction model curation. The first phase consists of perceptual nodes with limited computing resources that leads to the second phase, in which event data is worked through the routing nodes of the upper layer after data aggregation. Finally, in the last phase the central server produces an appropriate energy usage prediction. This prediction system is designed for enhanced decision support performance for energy planning, distribution, and conservation to reduce energy waste and environmental pollution. These costs are important to reduce and optimize since the multitude of IoT devices in a smart city require more complex electricity requirements, such as stable electric and back-up electric sources, to react in real-time. The final phase of the prediction system uses a deep neural network (DNN) due to their superior capacity to analyze extremely large quantities of data and exploit the spatiotemporal features of the data. This specific type of data is present in the sensor-heavy environments of smart cities, hence why DNNs have excelled in previous literature in comparison to simpler machine learning algorithms. Previous researchers have applied edge computing and deep learning for energy prediction models, however the novel application in this paper is using a combination of these fields to develop a model architecture capable of acquiring and processing data and regression prediction in real-time. For this model, the event data will be applied to many different scenarios and mapped into the prediction model. The paper concludes that edge computing architecture paired with deep learning algorithms can not only improve computational efficiency of short-term energy prediction by reducing central server load but can also do so in real-time.

As previously mentioned, smart cities dependent on many IoT devices and sensors generate mass amounts of data that need to be stored in a secured manner. [41] addresses this aspect of smart cities by proposing an end-to-end encryption process for edge-enabled smart cities called Smart Edge. Operating in a resource-constrained environment, [41] examines sensors called multimedia sensor networks (MSNs) to collect critical data streams from different applications within a smart city. The MSNs will then communicate with the cluster head for mutual authentication. If the mutual authentication is successful, a shared session key shared between the MSN and cluster head is generated for encrypted data transmission. Therefore, only authorized edge devices on a network can communicate to the cluster head for secured data transmissions. Figure 6 shows how the raw data is encrypted, sent to the cloud server, and decrypted by the end user. The storage of secured data from the edge and cloud service data centers is executed in 3 steps - that is encoding and encryption, partial decryption, and full decryption.

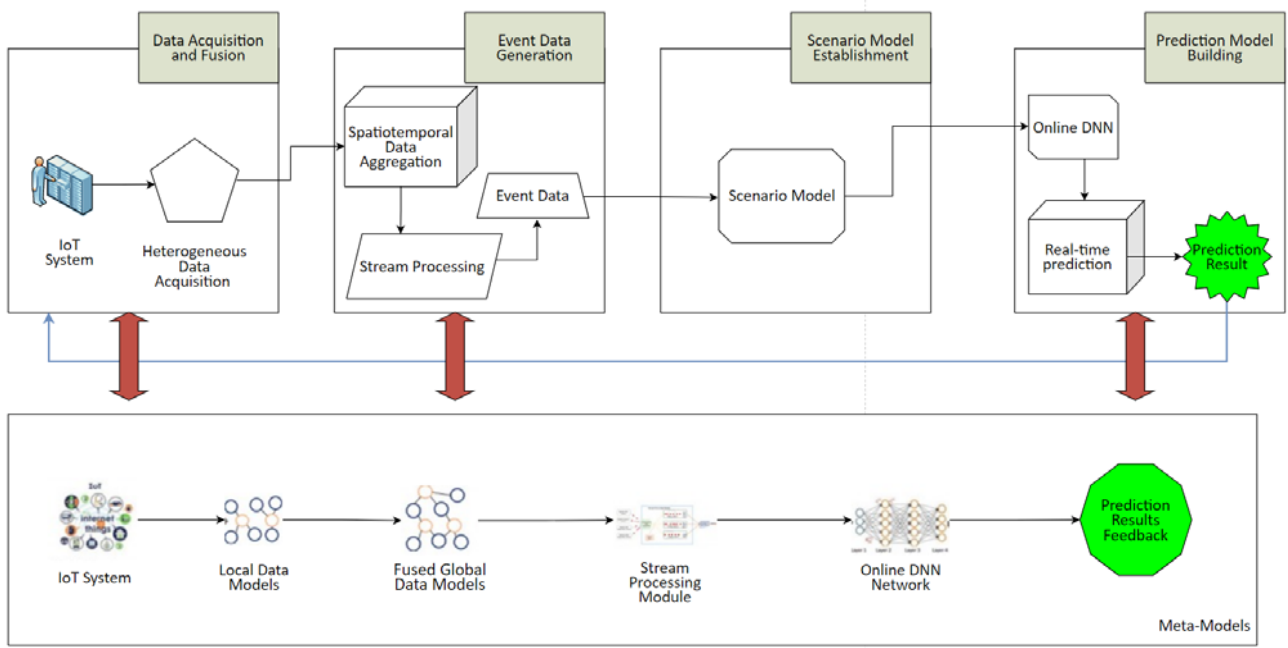

**Figure 5**. The framework of the short-term energy prediction system proposed in [40]

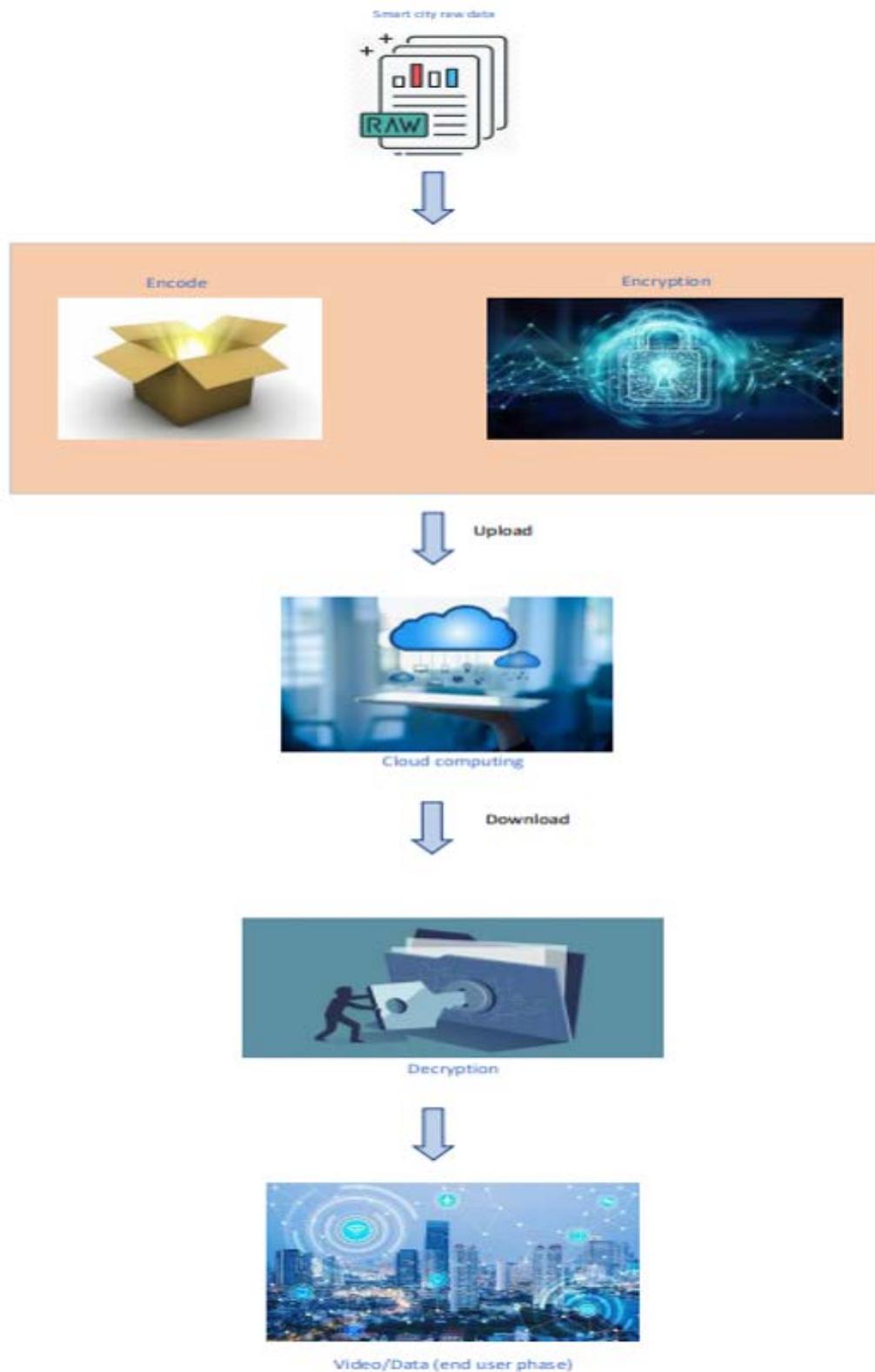

**Figure 6.** Secured data storage and sharing from [41]

The SmartEdge framework is used to compute the complex resource and power intensive operations between the edge devices and the cloud data center. The framework uses a lightweight symmetric encryption to make sure each multimedia sensor node (MSN) secures every data transmission with the cluster head. Due to the nature of symmetric encryption algorithms, prior to each data transmission, the edge computing processes must encrypt and encode the data stream before sending them through a secured connection to the cloud data centers for decryption with the same key. With this lightweight end-to-end encryption technique, there were a lowered number of session establishment delays as compared to the current alternative encryption methods. Despite its efficacy, the authors note that some future work includes configuring their scheme to incorporate mobility of IoT devices and increased scalability of the network edge devices.

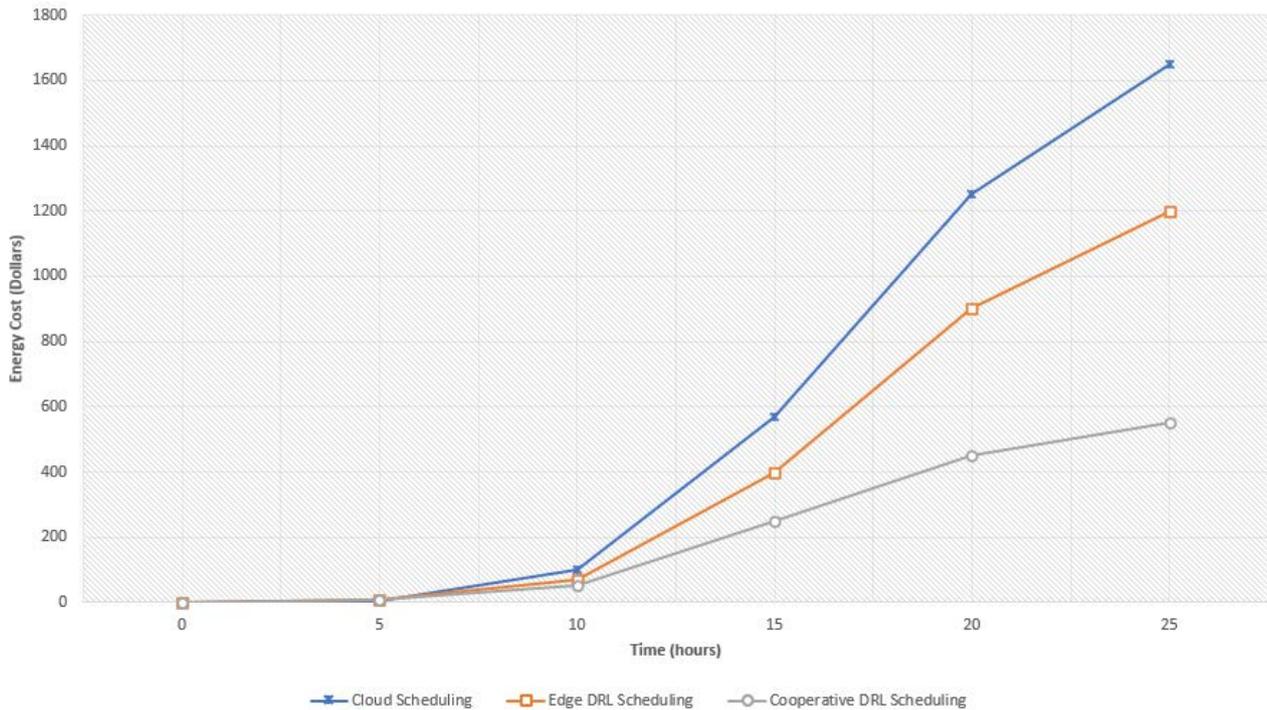

**Figure 7**. Energy cost comparison among different DRL methods. Taken from [42]

The significant advantage of saving energy through edge computing is again seen in [42] where the authors introduce an intelligent edge computing system to manage energy consumption in smart cities and homes. A novel innovation in this paper is the use of deep reinforcement learning (DRL) to improve long-term energy management performance as well as reduce standard execution times. This DRL scheme includes two phases: the use of an offline DNN to correlate value functions with corresponding actions and states, and the online deep Q-learning phase in which action selection, system control, and dynamic network updating is executed. Moreover, the authors evaluate the DRL model using two different model architectures: the edge DRL method and the cooperative DRL method. In the first method, the devices offload the energy scheduling task to the edge server to implement the DRL method for scheduling prediction optimization. The second method allows the edge server to further offload to a cloud server for DNN training and deep Q-learning process in order to curate a prediction from the resulting Q-value.

Figure 7 depicts how the cooperative DRL excels in reducing energy consumption when compared to the normal edge DRL and cloud scheduling. It is also noted in the paper that the expected energy cost reduction of all three scheduling methods increases proportionally with the number of homes. The authors conclude that this DRL-based energy management framework for a smart city is an improvement to traditional schemes and can be even further refined with future research.

### 3.3. Novel Edge Computing Techniques for Healthcare

Edge computing can also play an integral in the healthcare domain. Naturally, the healthcare system generates voluminous amounts of data by having to store records, medicine, previous visits, and medical history of each and every patient for an extended period of time. Thus, healthcare has become a prime candidate with which the performance of edge and cloud-based computing architectures has been tested against regular servers. Additionally, edge computing can assist in encrypting and securing patient data for reliably by offloading data processing from central servers. [43] explores these potential benefits by creating an edge computing-based framework for healthcare resource management called Resource Preservation Net (RPN). The task of RPN is to optimize in real-time the patient's length of stay, resource utilization rate, and average waiting time in the real-time, dynamic environment of a hospital's emergency room. It achieves this goal by using a graph optimization technique known as Petri net that has been proven to be reliable in analyzing workflow dynamics of a complex environment by exploiting graph theory. Thus, the RPN was tested in complex simulated environments, including but not limited to the healthcare domain to evaluate its ability to ensure limited resources are readily available as well as its general applicability. For example, Figure 8 exhibits how the RPN structure can optimize a patient's full stay through a hospital's emergency department.

This RPN was also applied to robotic automation to demonstrate the generality of the framework. In an industrial environment in which many items are being moved from one zone to another, heterogenous automated robots with distinct tasks must be able to operate amongst and with each other. The authors found that the complicated nature of synchronizing different robots meant for different tasks did not pose any difficulties for the RPN when creating an effective and time-efficient workflow during the simulated experiment.

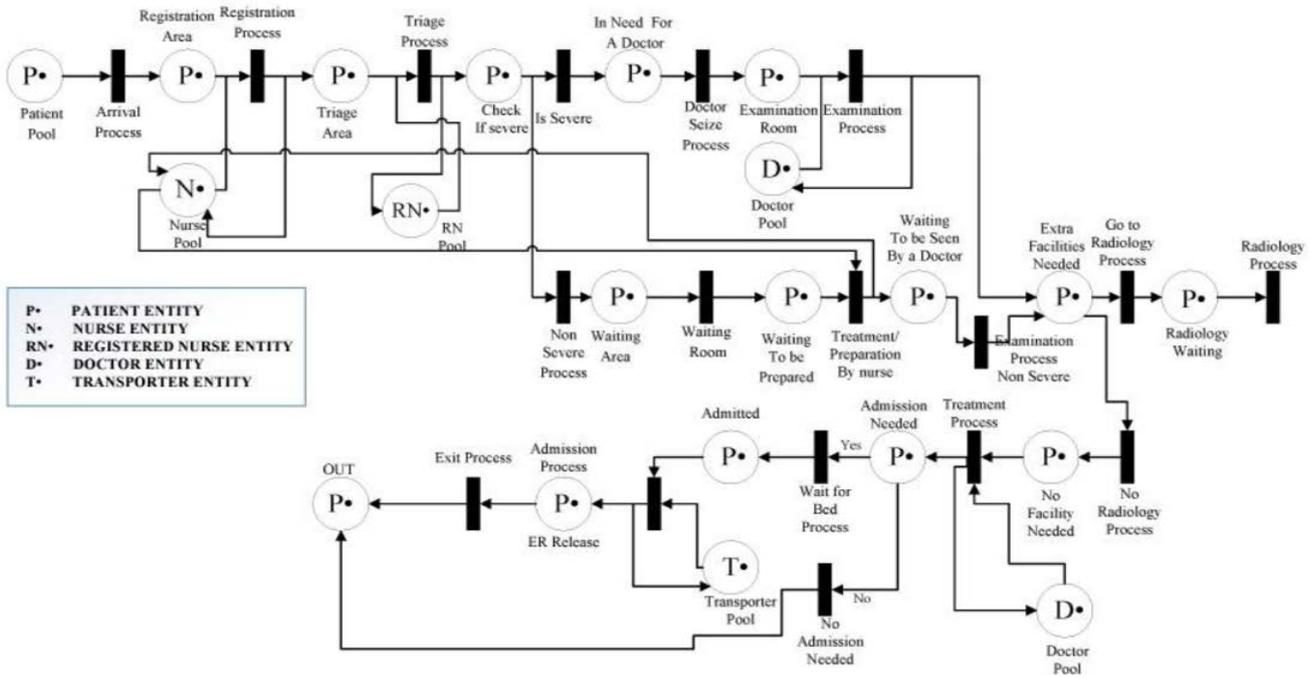

**Figure 8**. Simulated example of RPN application to healthcare environment from [43]

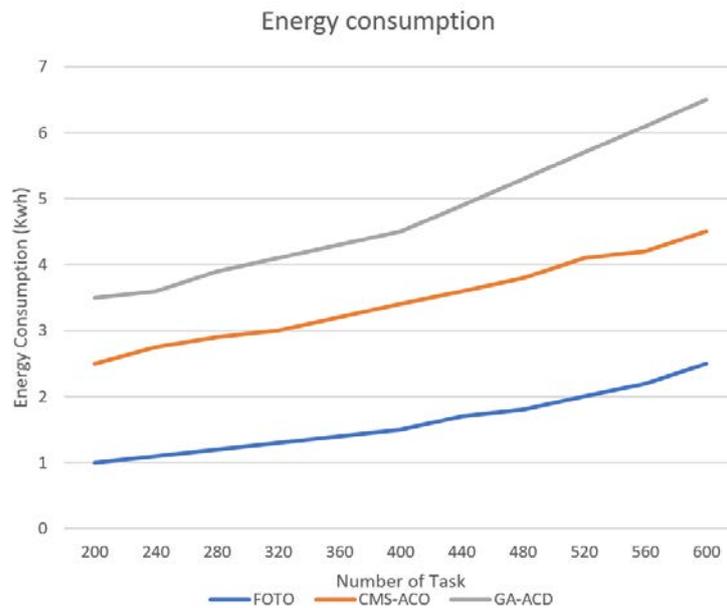

**Figure 9**. Energy Consumption diagram from [44]

Lastly, [44] introduces an alternative solution to limited resource scheduling in a smart healthcare system called fruit fly optimization-based task offloading algorithm (FOTO). The author describes the proper application of cloud resources to smart devices is an important process to optimize due to the low resource capacity in smart devices, thereby maximizing throughput of any healthcare application. The FOTO methodology is comprised of three parts: (i) Device manager, which optimizes and ensures edge node availability in the cloudlet (ii) Broker, which is responsible for the proper allocation of tasks to nodes in the cloudlet-based node availability, and (iii) Task tracker, which manages and completes the execution of each task in the nodes. A linear regression algorithm analyzes and is fitted to the incoming requests made by each user from the broker. The FOTO method can then be used to retrieve and assign the optimal amount of data to offload for a requested task, thereby reducing overall energy consumption and network congestion. FOTO was found not only to require less computational time, but in turn consumed less energy during operation when compared to alternative cooperative multi-task scheduling algorithms, namely the ant colony optimization algorithm (CMS-ACO) and heuristic queue-based algorithm (GA-ACO). Only 2.5 Kilowatt-hours (kWh) was consumed by FOTO when tested with 600 tasks while 4.3 kWh and 6.5 kWh was consumed by the other models, respectively. Figure 9 further demonstrates that FOTO can operate with lower energy consumption - and in turn, lower data center cost - even as the number of tasks increases.

# 4. Limitations in Current Literature

Across the reviewed literature, there are three limitations that are often mentioned with regards to edge computing. The first limitation being the limited implementation of 5G communication networks across most countries and cities. Edge computing usually perform best when operating in environments that are capable of producing sufficiently large amounts of data. Despite the implementation of 5G networks across major countries like China and the United States and the recent rapid growth seen across Europe and Asia, these networks still require additional time to become universally available to extract granular data more easily for edge computing. Some countries, like Sweden and Australia, have banned the 5G technology outright due to national security and infrastructure concerns. Thus, the sparse alternative environments that generate large quantities of data, like healthcare, will continue to be used for edge computing experiments until further 5G developments.

The second challenge resides in the model comparisons/optimizations and evaluation metrics. Because of relative infancy of edge computing, the varying computational capabilities of the many models tested in the reviewed literature are difficult to compare. Furthermore, machine and deep learning models are susceptible to converging to relatively local minima with regards to its functional analysis of loss. This makes it nearly impossible to fully optimize such large and complex models, especially in such a resource-constrained environments where edge computing is implemented. Additionally, evaluation metrics such as latency, bandwidth, energy usage, cost, etc. vary in usage across the literature. In order to concretely assert superior edge computing models for certain tasks over others, either universal metrics must be agreed upon or models must be tested in identical environments to mitigate bias.

Lastly, the concept of data abstraction is a big challenge for edge computing. To minimize the size of data, edge computing methods must extract only the most important features of the data to be sent to external computing clusters for use by the machine and deep learning algorithms. Edge computing, in its current state, is not currently equipped with the proper analysis techniques to extract important features from large datasets in real-time. In addition to this problem, some deep learning models - such as the DNNs used in the literature - build such abstract hierarchies from the data that important features are impossible to identify even after results have been disseminated. This abstraction of data can lead to low interpretability within the predictive models, making them difficult to optimize and improve upon, as seen in many limitations in Table 1.

**Table 1. General overview of literature review**

| Paper | Problem | Approach | Limitation |
|---|---|---|---|
| [35] | High latency and low spectral efficiency in centralized cloud computing | Propose three common edge computing technologies to mitigate deficiencies | Further research in edge computing needed to improve proposed models |
| [36] | High latency within current 5G networks and their supporting architectures | Combine existing NFV flexibility with vMEC intelligence to increase quality of service | Not enough IoT used in experiment to extrapolate results to large-scale networks |
| [37] | Edge computing and IoT has not been previously used to optimize factory production | Introduce a novel interconnected, intelligent robot production line using edge computing | Lacking in sufficiently reliable robotic intelligence, network load balancing, and intelligent scheduling methods |
| [38] | Centralized cloud computing faces performance problems caused by real-time remote network access | Use edge computing and SVMs reading EMR data to secure authentication between edge host and cloud server | Explore the possibility of applying deep learning algorithms for improved model accuracy |
| [39] | Smart cities and their cyberphysical systems need to adapt and grow with the physical environment and its demands | The iSapiens platform is responsible for the management of a distributed network of computing nodes to address geographical and functional extensibility | Implemented in one city in Italy. Different locations possess unique problems that iSapiens must be invariant to |
| [40] | The complexity and diversity of IoT data for energy management renders efficient energy prediction systems difficult to build | Propose a fully integrated model architecture in which edge computing is used to acquire and process energy data for a regression decision. | Lacking load balancing measures for larger-scale IoT experiments. DNN hyperparameters and general application can be further optimized. |
| [41] | High latency within data transmissions causes unacceptable delays in time-sensitive devices. Reducing this latency often causes highly unreliable data security during transmission. | Create a lightweight, symmetric, end-to-end encryption technique to securely transmit offloaded data from resource-constrained smart devices to network edge | Incorporate adaptability to mobile smart sensors and improve scalability of network edge. |
| [42] | Deep reinforcement learning has not been extensively applied as a solution to the voluminous problem space of efficient energy consumption in smart cities | Curate energy efficiency predictions in a game theory environment in which the reward for each DRL model is to achieve minimal energy consumption | Edge-DRL method relies too heavily on the limited computational capacity of edge servers for numerous devices' task. However, can be optimized to compete with the cooperative DRL method |
| [43] | Hospital emergency departments are complex, dynamic, real-time systems which contain limited resources – resources that must be optimized | Propose an RPN built on Petri net to effectively analyze the follow of a complex and concurrent system such as healthcare | Compliance with continually changing medical devices and privacy policies could prove difficult with the future of this work |
| [44] | Current methods of data offloading between nodes and networks in edge computing requires significant improvement | A generally applicable, flexible task offloading plan is proposed to obtain optimum resource allocation levels for requested jobs | This model has only been verified theoretically in simulations and has yet to provide significant results in a real-world experiment |

# 5. Conclusion

Through the review of previous literature, we have discussed the improvements edge computing can offer the research domains of IoT devices, smart cities, and smart healthcare systems. In addition, edge computing can reduce computational overhead and optimize data transmissions significantly such that some models are able operate in real-time. Most importantly, data transmissions across networks and edge devices are not only faster when optimized with edge computing techniques but were also seen to be both more secure and reliable.

For edge computing in relation to IoT devices, the proposed vMEC architecture seems to be one of the most effective and adaptable methods for optimizations. vMEC takes full advantage of growing 5G communications and has great configurability with currently existing network architectures, such as NFV. As mentioned in Section 4, this methodology is limited to only working in areas where 5G communication has been implemented. 5G is still a new technology and is slowly being rolled out in many different regions, but these issues may be resolved over time as countries continue to expand their infrastructure.

iSapiens is the most mature technology and has the potential to create the smart cities of the future using edge computing. It provided useful features for smart environments and physical systems to advance edge computing and IoT. Privacy problems may be a re-occurring issue with this platform, however, as consumers are becoming more hesitant with sharing their personal data with the public and private entities alike. This can be solved through strengthening the authentication methods within edge computing and providing more proof of concepts for the validity of future smart cities.

Regarding healthcare, RPN has the highest potential to be adopted into large-scale healthcare systems in order to increase overall quality of service to the patient. While previously proposed methods have included large, deep algorithms used to analyze patient data, RPN not only circumvents the necessity of powerful computing power to train and operate, but also does so in real-time. The speed at which data can be sent to be analyzed and returned to a healthcare professional can is quite important to an efficient medical operation, hence the need for increased research into edge computing in this topic.

Edge computing is the driving force of the improvements in fields like IoT, smart cities and smart healthcare systems. Therefore, if large breakthroughs are to be made in these fields, edge computing is an imperative generational technology that must receive continued support and analysis.